# En route to high T$_c$ superconductivity via Rb substitution of guest metal atoms in SrB$_3$C$_3$ clathrate

Peiyu Zhang[1], Xue Li[1], Xin Yang[1], Hui Wang[2], Hanyu Liu[1, 3, 4], Yansun Yao[5]

[1]*International Center for Computational Method & Software and State Key Laboratory of Superhard Materials, College of Physics, Jilin University, Changchun 130012, China*
[2]*Key Laboratory for Photonic and Electronic Bandgap Materials (Ministry of Education), School of Physics and Electronic Engineering, Harbin Normal University, Harbin 150025, China*
[3]*Key Laboratory of Physics and Technology for Advanced Batteries (Ministry of Education), International Center of Future Science, Jilin University, Changchun 130012, China*
[4]*International Center of Future Science, Jilin University, Changchun 130012, China*
[5]*Department of Physics and Engineering Physics, University of Saskatchewan, Saskatoon, Saskatchewan S7N 5E2, Canada*

**Abstract**

Recently, a host/guest clathrate SrB$_3$C$_3$ with *sp$^3$*-bonded boron-carbon framework was synthesized at around 50 GPa. On the basis of electron count, the structure is understood as guest Sr$^{2+}$ cations intercalated in the (B$_3$C$_3$)$^{3-}$ framework. Previous calculations suggest that SrB$_3$C$_3$ is a hole conductor with an estimated superconducting critical temperature (*T*$_c$) of 42 K at ambient pressure. If atoms with similar radius, such as Rb, can substitute Sr$^{2+}$ in the lattice, the electronic as well as superconductivity properties of this material will be modified significantly. Here, we perform extensive simulations on the stability and physical properties of Rb-Sr-B$_3$C$_3$ system using first-principles density functional calculation in combination with cluster expansion and CALYPSO structure prediction method. We predict a phonon-mediated superconductor Rb$_{0.5}$Sr$_{0.5}$B$_3$C$_3$ with a remarkably high *T*$_c$ of 78 K at ambient pressure, which is a significant improvement from the estimated value (42 K) in SrB$_3$C$_3$. The current results suggest that substitution of alkali atom in synthesized clathrate SrB$_3$C$_3$ is a viable route toward high-*T*$_c$ compounds.

**Introduction**

Ever since superconductivity was discovered in mercury at critical temperature ($T_c$) of 4.2 K in 1911 [1], research in superconductivity has been one of the central topics in Physics. Recently, most excitement in this field comes from hydrogen-rich compounds with $T_c$ above 200 K at high pressures [2-17], in which the superconductivity is still the Bardeen-Cooper-Schrieffer (BCS) [18] type and described by the Migdal-Eliashberg theory [19]. However, these high-$T_c$ hydrides require extremely high pressure to synthesize and usually not quench recoverable at ambient pressure, which limits their applications. Obviously, it is desirable to find new high-$T_c$ superconducting materials that are easy to synthesis and stable at ambient conditions.

Encouragingly, some non-hydrogen clathrates, in particular the $sp^3$ bonded ones with light elements, have been suggested to be superconducting at moderate or even ambient pressure. For example, several $p$-doped silicon clathrates based on $Ba_8Si_{46}$ have been synthesized below 3 GPa, but the highest observed $T_c$ is merely 8 K [20-23]. Many doped (especially $p$-type) $XC_6$ [24,25], $X(BN)_6$ [26], $XB_3C_3$ [27], and $XB_3Si_3$ [28] sodalite-like structures, where $X$ represents different doping elements, as well as $n$-doped $FC_{34}$ [29] have been predicted to be good superconductors at ambient pressure with the highest $T_c$ exceeding 100 K. However, most of the predicted materials have not been realized due to the experimental difficulties associated with thermodynamic instability. It is noteworthy that a thermodynamically stable host/guest clathrate $SrB_3C_3$ was synthesized at around 50 GPa [27]. This clathrate, with guest $Sr^{2+}$ cations incorporated at the voids in the $(B_3C_3)^{3-}$ host framework, is a hole conductor with the estimated $T_c$ of 42 K at ambient pressure [30]. A possible way to further enhance the electronic properties of this material is to replace $Sr^{2+}$ cations by alkali atoms, which would introduce more holes, but previous calculations shows that when all Sr atoms are replaced, the structure becomes dynamically unstable [31]. This leads us to consider the possibility to increase dynamic stability, while still bettering the properties, by a partial replacement of $Sr^{2+}$ cations in the clathrate. Preceding effort has been devoted to optimize the superconducting performance of hydrides through substitution or doping [32-38]; some already show significant enhancement in $T_c$ [32,36,37]. Given that Rb has similar atomic radius and electronegativity as Sr, it is suitable to be incorporated to the synthesized clathrate $SrB_3C_3$ to manipulate electronic properties and result in the significant modification of superconductivity.

Toward this end, we carried out a theoretical research on Rb-substituted $SrB_3C_3$. The compositional space of $Rb_xSr_{1-x}B_3C_3$ at 0 and 50 GPa was systematically explored using first-principles calculation in combination with cluster expansion (CE) [39] and CALYPSO structure prediction method [40-42]. Through this we found that the configurations with x ≤ 0.5 are dynamically stable, and unearthed several $Rb_xSr_{1-x}B_3C_3$ structures. Furthermore, the electron-phonon coupling (EPC) calculation and solution of the Eliashberg equation indicate that $Rb_{0.5}Sr_{0.5}B_3C_3$ in $Pm$-3 space group is a phonon-mediated superconductor with an estimated $T_c$ of 73-78 K at ambient pressure. These results provide guidance for future experimental studies and stimulate more exploration on boron-carbon-based high-$T_c$ superconducting materials.

**Computational details**

Stabilities of the $Rb_xSr_{1-x}B_3C_3$ alloys are investigated using the CE method [39], as implemented in the Massachusetts institute of technology *Ab initio* Phase Stability (MAPS) package in Alloy Theoretic Automated Toolkit (ATAT) [43]. In the CE method, alloy is treated by the lattice model where the lattice sites are fixed on the Bravais lattice and a configuration σ is defined by specifying the occupation of each lattice site. For quasibinary $RbB_3C_3$-$SrB_3C_3$ system, the alloy configuration is described by pseudospin occupation variable $\sigma_i$, which takes the value -1 and +1 when the *i*th site is occupied by Rb and Sr, respectively. The energy (or enthalpy) for a given configuration σ is expressed by the following polynomial:

$$E(\sigma) = \sum_\alpha m_\alpha J_\alpha \left\langle \prod_{i \in \alpha'} \sigma_i \right\rangle,$$

where $\alpha$ is a cluster and $J_\alpha$ is the effective cluster interaction (ECI). The multiplicity factor $m_\alpha$ indicates the number of equivalent clusters to $\alpha$ by symmetry. The precision of the ECIs can be evaluated by the cross-validation (CV) score, defined as:

$$CV = \frac{1}{n}\sum_{\alpha}^{n}(E_i - \hat{E}_{(i)})^2,$$

where *n* is the number of structures used to obtain the ECIs. $E_i$ is the energy of the *i*th structure calculated using density functional theory (DFT) and $\hat{E}_{(i)}$ is the predicted energy of the same structure obtained from CE method fitting to the energies of remaining (N-1) structures. To ensure a satisfactory convergence of CEs, the CV score should be as small as possible. In the present work, a cubic *Pm*-3*n* structure of $SrB_3C_3$ is used as the parent structure, and the CV scores are 2.2 and 2.5 meV/atom at 0 and 50 GPa, respectively.

The DFT calculations for fitting the ECI were performed with the Vienna Ab Initio Simulation Package (VASP) [44], using projector augmented waves (PAW) pseudopotentials [45] with Perdew-Burke-Ernzerhof (PBE) generalized gradient approximation (GGA) exchange-correlation functional [46]. A plane-wave energy cutoff of 500 eV and a Monkhorst-Pack Brillouin Zone (BZ) sampling grid of $2\pi \times 0.032$ Å$^{-1}$ were employed. The convergence criteria were set to $10^{-5}$ eV per unit cell for energy and 0.02 eV/Å per atom for forces.

To examine the validity of the CE method, we performed a crystal structure search for $RbSrB_6C_6$ using a heuristic algorithm based on the particle swarm-intelligence approach as implemented in the CALYPSO code [40-42]. The structures of $RbSrB_6C_6$ were searched using simulation cells with 1-2 formula units (f.u.) at 50 GPa. The total number of predicted structures is at least ~2000.

To determine the dynamic stabilities of predicted structures by the CE method, phonon calculations were performed using the finite displacement method implemented in the PHONOPY software [47]. The EPC calculations for superconducting properties were performed with density functional perturbation theory (DFPT) [48] using Quantum ESPRESSO package [49]. GBRV ultrasoft pseudopotentials [50] were used for Rb, Sr, B and C, with a kinetic energy cutoff of 60 Ry, a dense *k*-point mesh with a spacing of $2\pi \times 0.008$ Å$^{-1}$, and a *q*-point mesh with a spacing of $2\pi \times 0.034$ Å$^{-1}$. The superconducting gap and $T_c$ were obtained by numerically solving the Eliashberg equation [19]. This equation provides a reasonable estimate of $T_c$ for strong-coupling cases with EPC parameter $\lambda$ above 1.5.

**Results and discussions**

The CE simulations were performed on crystal structures with up to 84 atoms. To ensure a satisfactory convergence, 141 and 113 structures were used for fitting while 6342 and 6370 structures were employed for prediction at 0 and 50 GPa, respectively. The corresponding CV scores are 2.2 and 2.5 meV/atom. Fig. 1 shows the calculated formation enthalpies for various $Rb_xSr_{1-x}B_3C_3$ compositions at 0 and 50 GPa. We found 14 and 11 stable structures of $Rb_xSr_{1-x}B_3C_3$ at 0 and 50 GPa, respectively. Among them, a cubic *Pm*-3 structure of $Rb_{0.5}Sr_{0.5}B_3C_3$ has the lowest energy at both 0 and 50 GPa. In addition, we performed a CALYPSO structure search for fixed $Rb_{0.5}Sr_{0.5}B_3C_3$ composition and found the same *Pm*-3 as the stable phase, indicating the present computational scheme is reliable. Phonon calculations were carried out to investigate the dynamical stability of the predicted structures. The results show that $Rb_xSr_{1-x}B_3C_3$ structures with x≤0.5 do not have imaginary frequencies in the whole BZ at both 0 and 50 GPa (Fig. S1). Therefore, we select candidate structures with x≤0.5 for further analysis and discussion.

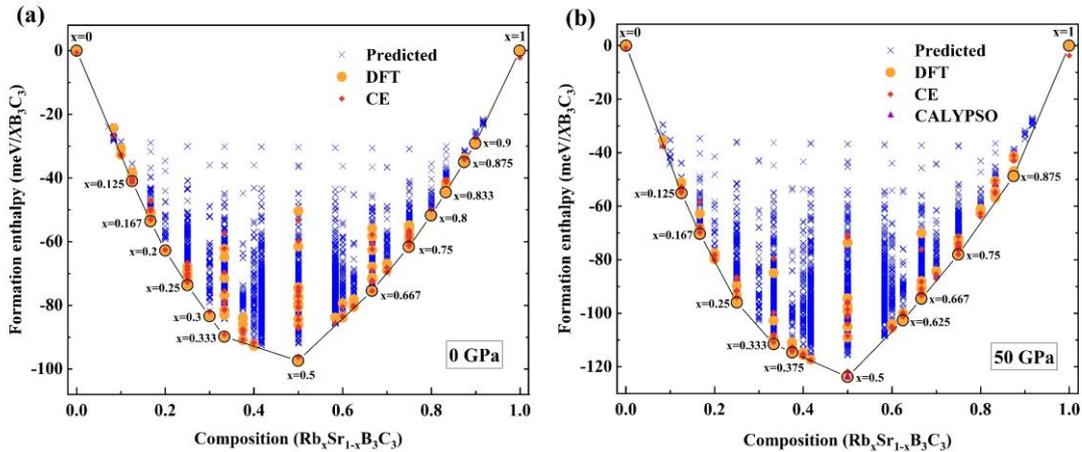

Fig. 1. Formation enthalpies (in unit of meV/$X$B$_3$C$_3$) and corresponding convex hull of the $Rb_xSr_{1-x}B_3C_3$ system at (a) 0 and (b) 50 GPa. The two end points are ternary SrB$_3$C$_3$ and RbB$_3$C$_3$,

both in *Pm*-3*n* structure. Blue crosses denote structures predicted from the CE method. Solid orange circles correspond to the enthalpies of structures computed with DFT, while the red squares represent the enthalpies from the corresponding CE. Purple triangles indicate the ground state structures predicted by the CALYPSO code.

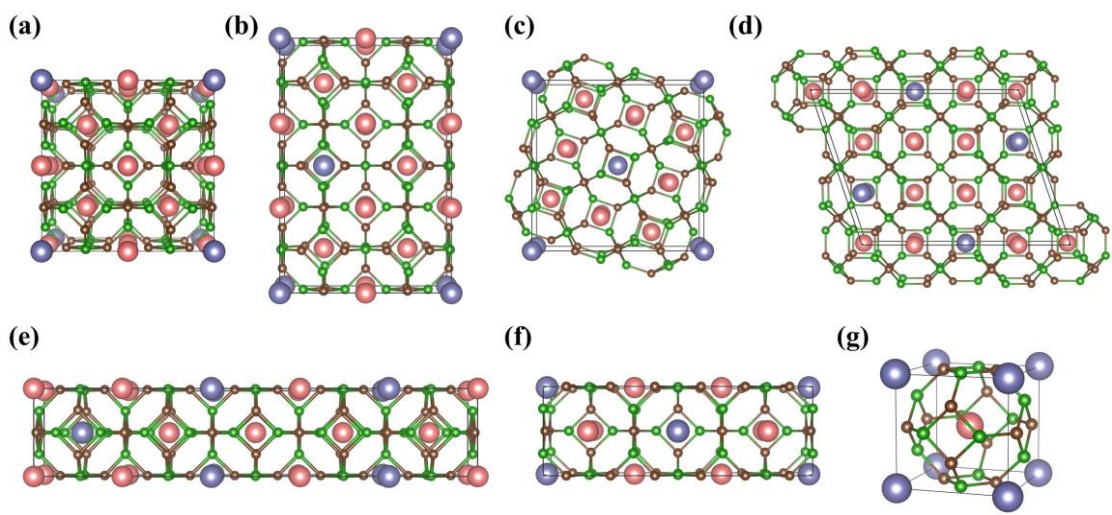

Fig. 2. Predicted structures of $Rb_xSr_{1-x}B_3C_3$ for (a) $x$ = 0.125, (b) 0.167, (c) 0.2, (d) 0.25, (e) 0.3, (f) 0.333, and (g) 0.5 at 0 GPa. The Rb, Sr, B and C atoms are shown in purple, pink, green and brown, respectively. Detailed information of predicted structures is summarized in Tables S1 and S2 in the Supplemental Material.

All predicted structures, as shown in Fig. 2, can be viewed as host-guest structures. The B and C atoms form interconnected host framework, and the guest metals occupy the centers of the cubic units with different Rb/Sr ratios. The B-C cage is composed of 24 alternatively arranged B and C atoms, of which eight $B_3C_3$ hexagonal faces and six $B_2C_2$ rhombic faces are connected to each other by sharing B-C edges to form a $B_6C_6$ sodalite-like cage. Each cage accommodates one of the Rb/Sr atoms in the center. Due to the intrinsic differences between Rb and Sr atoms, slight distortions occur in the B-C cages for the doped compounds. In these cages, the B-C bond lengths range between 1.71 and 1.75 Å, and C-B-C (B-C-B) angles in between 88° and 92° (119° and 121°). They differ slightly from the parent structure $SrB_3C_3$, which has a unique bond length (1.73 Å) and angles (90° and 120°) [27] at ambient pressure.

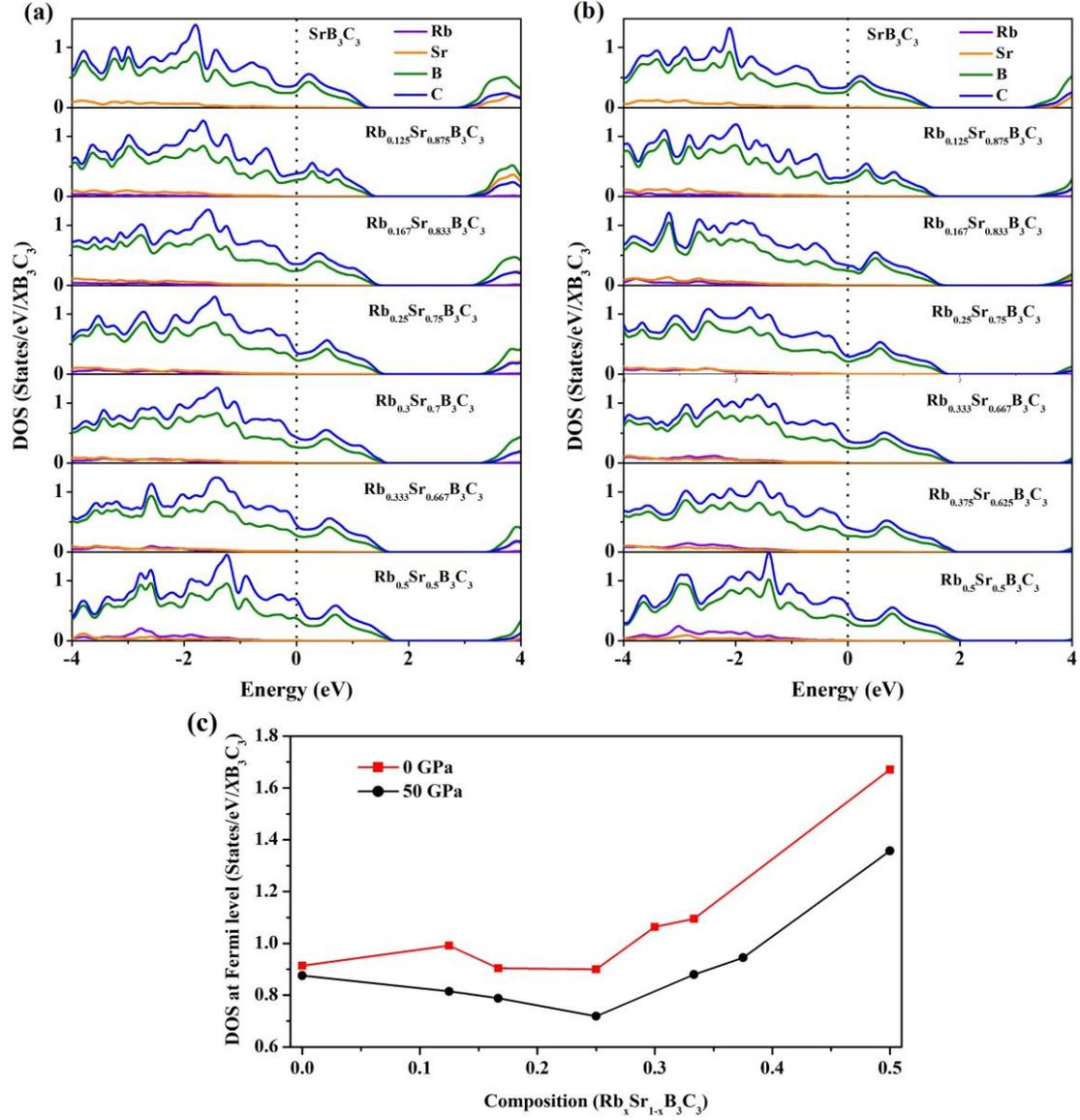

Fig. 3. Calculated PDOS for various Rb contents in $Rb_xSr_{1-x}B_3C_3$ at (a) 0 and (b) 50 GPa. Vertical dashed lines indicate the Fermi level. (c) Variation of total DOS at the Fermi level with composition.

The electronic properties of the $Rb_xSr_{1-x}B_3C_3$ at different Rb concentrations are revealed by the projected electronic density of states (PDOS), as shown in Figs. 3(a) and 3(b). The results show that the electrons near the Fermi level are mainly from B and C atoms (cages), while the contribution from Rb and Sr atoms are negligible. The evolution of total DOS at the Fermi level with substitution concentration $x$ at 0 and 50 GPa are displayed in Fig. 3(c). Obviously, the DOS at the Fermi level decreases first and then increases as the Rb substitution rate increases. It is noteworthy that once the Rb content reaches 0.5, the DOS increases sharply to 1.67 and 1.36 (in units of States/eV/$X$B$_3$C$_3$) at 0 and 50 GPa, respectively, which is much higher than the value of ~0.9 in SrB$_3$C$_3$. This is because Rb doping changes the number of electrons transferred from the metal elements to the B-C host framework, and therefore the Fermi level shifted down accordingly [see Figs. 3(a) and 3(b)]. A high value of DOS at the Fermi level is critical to raising the possibility of the formation of Cooper pairs in Rb$_{0.5}$Sr$_{0.5}$B$_3$C$_3$, thus enhancing electron-phonon coupling strength and superconductivity.

To investigate potential superconductivity in Rb$_{0.5}$Sr$_{0.5}$B$_3$C$_3$, we performed the EPC calculation and estimated the superconducting $T_c$ through solving the Eliashberg equation [19]. Typical Coulomb pseudopotential parameters $\mu^*$ from 0.1 to 0.13 are used in the solution. Fig. 4 shows the calculated phonon dispersions, projected phonon density of states (PHDOS) to individual atoms, Eliashberg spectral function $\alpha^2F(\omega)/\omega$, and integrated EPC strength $\lambda$ for Rb$_{0.5}$Sr$_{0.5}$B$_3$C$_3$ at 0 GPa. Obviously, coupling of electrons to phonons in Rb$_{0.5}$Sr$_{0.5}$B$_3$C$_3$ mainly originates from high frequency vibrations in the range of 200-800 cm$^{-1}$, which contribute 69% to the total $\lambda$. Intriguingly, these frequencies are dominated by the vibrations of B-C framework. The resulting $T_c$ reaches a maximum value of 78 K using $\mu^* = 0.1$ at 0 GPa, which is comparable to the liquid nitrogen temperature (77 K). Furthermore, the effects of pressure on the $T_c$ were examined for Rb$_{0.5}$Sr$_{0.5}$B$_3$C$_3$. The results show that $T_c$ has a significant drop to 47 K at 50 GPa, driven by crystal lattice hardening and resulting reduction of $\lambda$ from 2.06 at 0 GPa to 0.92 at 50 GPa (See Fig. S2 in the Supplemental Material for results at 50 GPa for comparison with the data shown in Fig. 4.) As well, the superconducting gap obtained from solving Eliashberg equation also shows a notable drop at high pressure [Fig. 5(a)].

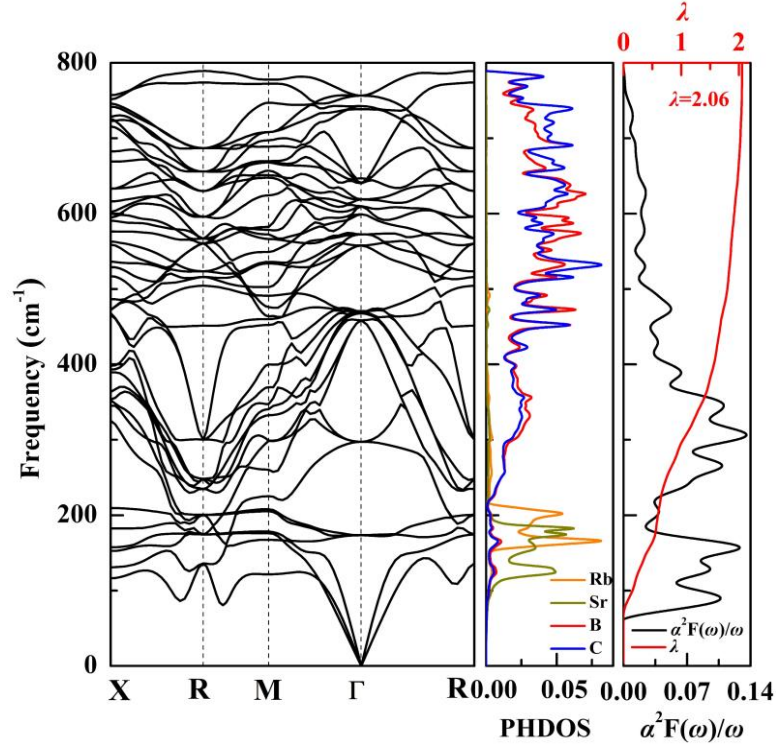

Fig. 4. Calculated phonon dispersion relations, PHDOS, and Eliashberg spectral function of $Rb_{0.5}Sr_{0.5}B_3C_3$ at 0 GPa.

We have also evaluated the $T_c$s for other structures with different Rb content, such as $SrB_3C_3$ ($Rb_0SrB_3C_3$) and $Rb_{0.333}Sr_{0.667}B_3C_3$. The predicted $T_c$ of $SrB_3C_3$ is consistent with previously published theoretical results [30]. $T_c$s are estimated to be 46 K and 69 K at 0 GPa for $SrB_3C_3$ and $Rb_{0.333}Sr_{0.667}B_3C_3$, respectively. These two species have lower $T_c$ mainly because of a lower $\lambda$ and $N_{Ef}$ compared to $Rb_{0.5}Sr_{0.5}B_3C_3$, as shown in Fig. 5(b). In addition, we have considered other 24 compounds of $X_{0.5}Y_{0.5}B_3C_3$ ($X$ = Li, Na, K, Rb, Cs; $Y$ = Be, Mg, Ca, Sr, Ba), which are isoelectronic as well as isostructural to $Pm$-3-$Rb_{0.5}Sr_{0.5}B_3C_3$. Phonon calculations show that only $K_{0.5}Ca_{0.5}B_3C_3$ and $K_{0.5}Sr_{0.5}B_3C_3$ are dynamically stable at ambient pressure. We have subsequently explored the superconductivity of these structures, in comparison with $Rb_{0.5}Sr_{0.5}B_3C_3$ [Fig. 5(b)]. According to the isotope effect, system with a lower mass is likely to have a higher $T_c$. However, $T_c$s of both $K_{0.5}Ca_{0.5}B_3C_3$ (76 K) and $K_{0.5}Sr_{0.5}B_3C_3$ (75 K) do not exceed the $T_c$ of $Rb_{0.5}Sr_{0.5}B_3C_3$ at ambient pressure. A greater contribution of high frequency phonons to the superconductivity is offset by the decrease in EPC strength $\lambda$, resulting in a comparable $T_c$ in $K_{0.5}Ca_{0.5}B_3C_3$ and $K_{0.5}Sr_{0.5}B_3C_3$.

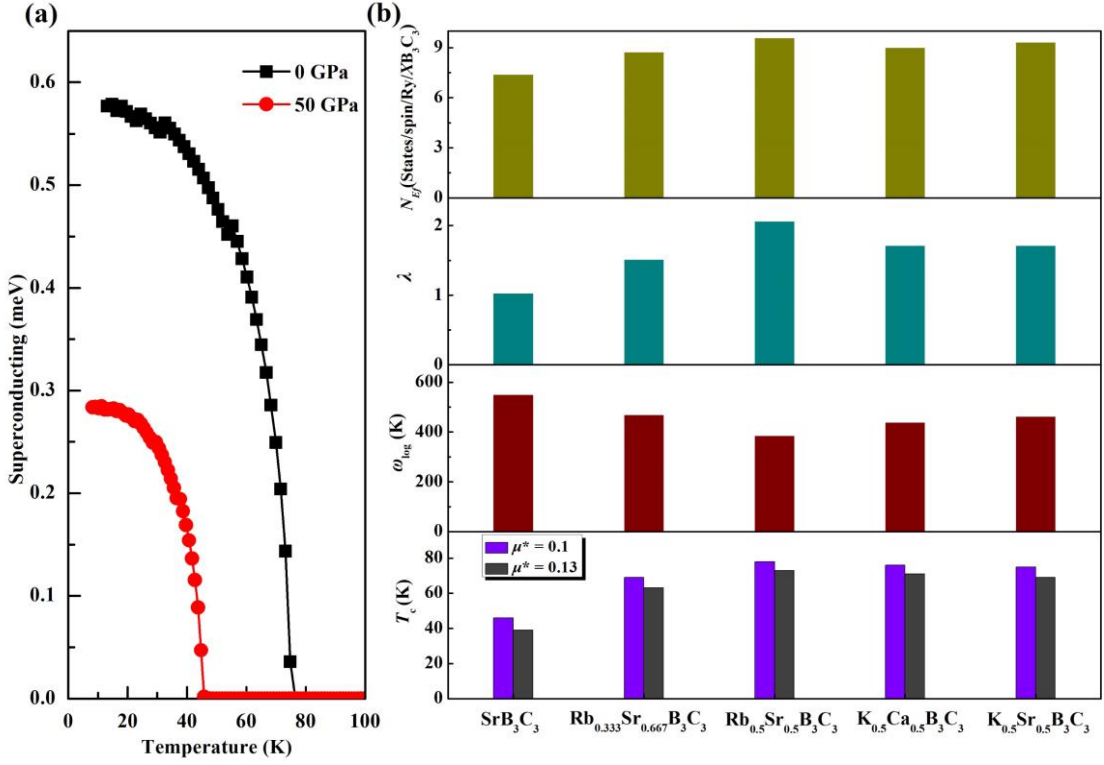

Fig. 5. (a) Superconducting gap of $Rb_{0.5}Sr_{0.5}B_3C_3$ calculated at 0 and 50 GPa. (b) Top to bottom: electronic DOS at the Fermi level ($N_{Ef}$), EPC parameter $\lambda$, logarithmic average phonon frequency $\omega_{\log}$, and $T_c$ of various clathrate structures at 0 GPa.

In order to investigate the thermodynamic stability of $Rb_{0.5}Sr_{0.5}B_3C_3$, we systematically examined the formation enthalpies as a function of pressure. Fig. 6 shows the enthalpy differences in several selected reaction paths in the pressure range of 0-120 GPa. It is clearly seen that the $Rb_{0.5}Sr_{0.5}B_3C_3$ phase has higher enthalpy than some combinations of elements, binary or ternary compounds, indicating that it is a metastable phase. Encouragingly, the results show that the $Rb_{0.5}Sr_{0.5}B_3C_3$ phase would be stable compared with $RbB_6$, $SrC_2$ and C above 60 GPa, suggesting that this material can be synthesized for this special reaction route under extreme pressure. The dynamics stability of this phase suggests that it is quenchable to ambient conditions, following the same thermodynamic route as the parent $SrB_3C_3$.

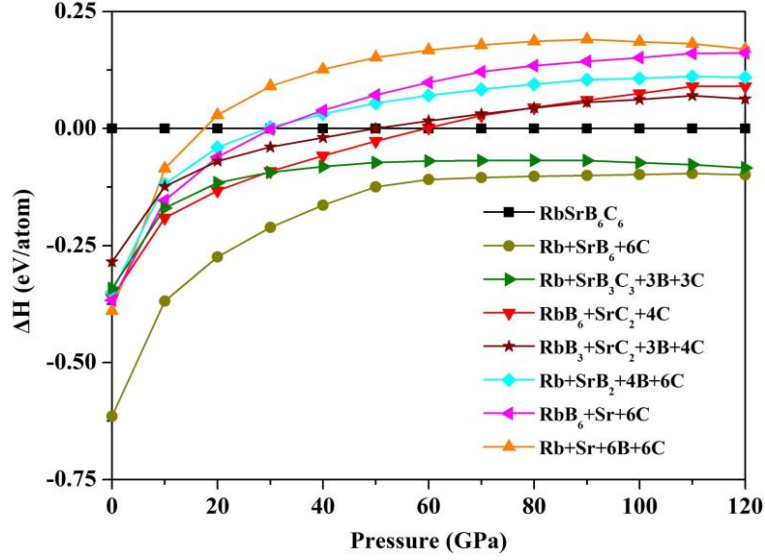

Fig. 6. Calculated enthalpy for $Rb_{0.5}Sr_{0.5}B_3C_3$ as a function of pressure.

**Conclusions**

In summary, we have systematically explored the phase stability of the complete compositional space of $Rb_xSr_{1-x}B_3C_3$ at 0 and 50 GPa using first-principles methods in combination with cluster expansion and CALYPSO structure prediction methods. A number of $Rb_xSr_{1-x}B_3C_3$ structures were discovered, among which the structures with x≤0.5 are dynamically stable. Electron phonon coupling calculations indicate that *Pm-3*-$Rb_{0.5}Sr_{0.5}B_3C_3$ phase is a promising conventional superconductor with estimated $T_c$ of 78 K at ambient pressure. In addition, comparable $T_c$ values were also predicted in $K_{0.5}Ca_{0.5}B_3C_3$ and $K_{0.5}Sr_{0.5}B_3C_3$ compounds, where $T_c$ reaches 76 and 75 K, respectively. Our findings advance the understanding of superconducting materials consisting of low-Z elements, such as boron and carbon, under ambient conditions and stimulates future theoretical and experimental studies in this field.

**Acknowledgments**

This work is supported by the National Natural Science Foundation of China (Grants No. 52090024, 12074138, 11974135, 11874175, and 11874176), the Fundamental Research Funds for the Central Universities (Jilin University, JLU), the Program for Jilin University Science and Technology Innovative Research Team (JLUSTIRT), and the Strategic Priority Research Program of Chinese Academy of Sciences (Grant No. XDB33000000). We used the computing facilities at the High-Performance Computing Centre of Jilin University.

Corresponding authors: Hanyu Liu(hanyuliu@jlu.edu.cn) and Yansun Yao(yansun.yao@usask.ca)